\documentclass[journal]{IEEEtran}

\usepackage{amsmath}
\usepackage{amsfonts}
\usepackage{graphicx}
\usepackage{booktabs}
\usepackage{array}
\usepackage{cite}
\usepackage{url}
\usepackage{dblfloatfix}  % allows figure* at bottom of page in two-column

\begin{document}

\title{Stabilizing Illumination Conditioning in SSM-Based Low-Light Image Enhancement with Bottleneck Prompt Sidecars}

\author{Wei~Chen~Ng,~
        Qiyi~Xie,~
        Jian~Ding~Tan$^{*}$,~
        Yan~Kang~Lee,~
        Abdulrab~Hakim~Qaid~Abdullah,~
        and~~Wei~Hown~Tee%
\thanks{$^{*}$Corresponding author: Jian Ding Tan (e-mail: jianding.tan@xmu.edu.my).}%
\thanks{W.~C.~Ng is with the School of Computing and Data Science, Xiamen University Malaysia, Sepang, Selangor 43900, Malaysia.}%
\thanks{Q.~Xie is with the School of Artificial Intelligence and Robotics, Xiamen University Malaysia, and with the School of Engineering and Applied Science, University of Pennsylvania, Philadelphia, PA 19104, USA.}%
\thanks{J.~D.~Tan and A.~H.~Q.~Abdullah are with the School of Artificial Intelligence and Robotics and the S.M.A.R.T. Nexus Center of Excellence, Xiamen University Malaysia, Sepang, Selangor 43900, Malaysia.}%
\thanks{Y.~K.~Lee is with the School of Energy and Chemical Engineering, Xiamen University Malaysia, Sepang, Selangor 43900, Malaysia.}%
\thanks{W.~H.~Tee is with the School of Engineering, Taylor's University, 47500 Subang Jaya, Selangor, Malaysia.}%
\thanks{This work was supported by the Xiamen University Malaysia Research Fund under Grant XMUMRF/2022-C10/IECE/0040.}}

\markboth{Preprint}{Ng \MakeLowercase{\textit{et al.}}: Stabilizing Illumination Conditioning in SSM-Based LLIE}

\maketitle

\begin{abstract}
Low-light image enhancement (LLIE) requires more than brightness amplification because real low-light images often exhibit spatially varying illumination, amplified noise, color distortion, and local detail loss. Recent SSM- and Mamba-based restorers provide efficient long-range modeling, yet the route for introducing illumination priors into such backbones remains insufficiently studied. This study addresses safe illumination conditioning for SSM-based LLIE through a bottleneck-side Prompt-Adapter. A lightweight illumination estimator first predicts a single-channel illumination map, which is then converted into prompt tokens and injected through a residual sidecar adapter at the bottleneck, thereby avoiding direct modification of the state-space core. Under a locked three-seed evaluation protocol on LOL-v1, LOLv2-Real, and LOLv2-Synthetic, the Prompt-Adapter improves over the Base model by $0.94$ dB, $1.24$ dB, and $0.93$ dB in PSNR, respectively, while also improving SSIM, LPIPS, and masked dark-region metrics. The clearest real-data evidence appears on LOLv2-Real. A Learned-Fixed control on LOL-v1 indicates that the present results support the utility of the sidecar conditioning pathway more strongly than a strict claim of illumination-conditioned superiority, and the LOLv2-Synthetic result remains stability-conditioned. These findings suggest that, in SSM-based LLIE, conditioning design can be as important as the conditioning signal itself.
\end{abstract}

\renewcommand{\IEEEkeywordsname}{Keywords}

\section{Introduction}

\IEEEPARstart{L}{ow-light} image enhancement (LLIE) is a fundamental image restoration problem with practical importance in photography, surveillance, robotics, and other vision systems that operate under limited illumination. The difficulty of LLIE does not arise from insufficient brightness alone. Real low-light images usually involve spatially uneven illumination, noise amplification, color deviation, and weakened local structure, so na\"{\i}ve brightening often improves visibility at the cost of fidelity. A useful LLIE method must therefore recover image content while preserving the illumination-dependent structure of the degradation.

Existing LLIE methods approach this problem from several directions. Traditional and Retinex-based models emphasize illumination decomposition and prior-guided enhancement~\cite{singh2025lliereview,sun2025diretinex}, while more recent deep models strengthen restoration capacity through Transformer, generative, and state-space architectures~\cite{zhou2025luxformer,xie2026igdbt,xu2025mambahvi,liu2025multistreamvssm}. These developments have improved the quality of low-light restoration, but they leave a narrower methodological question unresolved. In an SSM-based LLIE model, the central issue is not only whether illumination information is useful, but also how that information should be introduced into the restoration backbone.

This distinction matters because an informative prior can still be harmful if the conditioning route is poorly chosen. In preliminary direct-injection probes, directly coupling illumination-aware signals with the state-space core did not yield a stable gain and instead introduced unstable or visually unsafe behavior. The resulting design shift is central to the present study. Rather than treating illumination guidance as a generic auxiliary cue or pursuing a larger backbone redesign, the study focuses on safe illumination conditioning for SSM-based LLIE.

The final method adopts a bottleneck-side Prompt-Adapter route. A lightweight illumination estimator first predicts an illumination map from the low-light input. This signal is then converted into prompt tokens and injected through a residual sidecar adapter at the bottleneck of the restorer, thereby preserving the main backbone and avoiding direct modification of the state-space core. Under a locked three-seed protocol, this route improves over the Base model on LOL-v1, LOLv2-Real, and LOLv2-Synthetic. The strongest real-data evidence appears on LOLv2-Real. At the same time, the interpretation remains deliberately bounded: the Learned-Fixed control does not support a strict claim that illumination-conditioned prompts are superior to fixed learned prompts in mean performance, and the LOLv2-Synthetic result retains a stability caveat.

Relevant prior work can therefore be read through three themes: illumination-aware enhancement, zero-reference and lightweight restoration, and global-modeling architectures such as Transformer, diffusion, and state-space based LLIE. These themes provide the immediate context for the present study.

\subsection{Related Work}

\paragraph{Illumination-aware and Retinex-based LLIE}
Illumination modeling remains one of the dominant design principles in low-light image enhancement. A recent review by Singh and Bhandari~\cite{singh2025lliereview} organizes this literature into traditional enhancement, Retinex-based modeling, zero-reference learning, supervised deep restoration, and recent generative frameworks. Within this line, DI-Retinex revisits the Retinex formulation from the perspective of digital imaging and shows that practical LLIE must account for imaging factors that are often ignored in simplified decomposition pipelines~\cite{sun2025diretinex}. More recent models retain this illumination-aware perspective while strengthening feature modeling. LUXFormer performs joint spatial-frequency illumination modeling~\cite{zhou2025luxformer}, while dual-guide and reflectance-oriented generative models introduce illumination guidance through normalizing flow, latent diffusion, or hybrid reflectance reconstruction~\cite{zhuang2026dualguideflow,kolekar2026rldtnet,yang2026reflectanceflow}. Other methods further combine bilateral priors or coordinated enhancement objectives to balance illumination correction with detail preservation~\cite{lv2026bipcenet,cao2026nidllie}. These studies strongly support the value of illumination priors, but they usually redesign the enhancement pipeline around decomposition, guided branches, or full generative reconstruction. The question of how to inject illumination information into an SSM-based restorer without destabilizing the backbone remains less explicit.

\paragraph{Zero-reference and lightweight LLIE}
Another major branch seeks to reduce supervision requirements or computational cost. Wu \emph{et al.} analyze the mechanism of zero-reference deep curve estimation and show that the effectiveness of this route depends strongly on how image-specific curve adjustment is parameterized and regularized~\cite{wu2024zerodceanalysis}. Recent methods extend this idea through multi-scale curve estimation, chroma-aware implicit representations, and compact wavelet-domain enhancement networks~\cite{wang2026mpyace,wang2026cdin,an2026lwinet}. Efficiency-oriented designs such as TOVO and NID-LLIE further emphasize controllable enhancement under lightweight constraints~\cite{razafindratovolahy2026tovo,cao2026nidllie}, while HDD-Unet shows that even supervised compact architectures still focus primarily on restoration efficiency and image-quality preservation rather than on structured conditioning design~\cite{batziou2026hddunet}. This literature highlights an important trade-off between enhancement capacity and practical controllability, but it still focuses on the mapping function itself more than on the interface between an illumination prior and a structured global-modeling backbone.

\paragraph{Transformer, diffusion, and state-space LLIE}
Recent LLIE methods increasingly rely on architectures with stronger global reasoning capacity. Transformer-based methods have been used to capture long-range dependencies through multi-stage, dual-branch, or iterative attention designs~\cite{tan2026msoctransformer,xie2026igdbt,zhang2026sianet}. At the same time, state-space modeling has started to enter LLIE. Selective state-space models such as Mamba~\cite{gu2023mamba} provide input-dependent long-range modeling with linear complexity in sequence length, and MambaIR~\cite{guo2024mambair} established a simple state-space baseline for general image restoration. For LLIE specifically, MambaLLIE~\cite{weng2024mamballie} combines a global-then-local state-space block with an implicit Retinex-aware selective kernel module, and RetinexMamba~\cite{bai2024retinexmamba} couples Retinex decomposition with a Mamba backbone. Xu and Hao adopt an HVI-based Mamba-transformer design~\cite{xu2025mambahvi}, while Liu \emph{et al.} introduce a multi-stream vision state space module for low-light restoration~\cite{liu2025multistreamvssm}. In parallel, generative modeling has become increasingly prominent, with pretrained CNN-guided diffusion, probabilistic reasoning, and data-synthesis driven diffusion priors being used to improve robustness under complex degradations~\cite{tu2026pretraineddiffusion,wei2026counterfactualllie,wang2026wilddiffusionprior}. These studies indicate that LLIE benefits from stronger priors and richer global modeling. However, they mostly study new end-to-end restoration architectures rather than the narrower conditioning problem considered here. The present study therefore differs in emphasis: instead of treating LLIE progress as a sequence of larger architectural redesigns, it examines how an illumination prior can be introduced into an SSM-based restorer through a safer sidecar pathway rather than through direct modification of the state-space core.

\subsection{Contributions}

The contributions are summarized as follows.
\begin{enumerate}
\item A safe-conditioning perspective is introduced for SSM-based LLIE, framing the problem as one of how illumination priors should enter the restoration backbone rather than as a generic architecture-stacking exercise.
\item A bottleneck-side Prompt-Adapter pathway is designed to inject illumination-aware side information through a residual sidecar route without directly rewriting the state-space core dynamics.
\item The proposed route is evaluated under a locked and repeated protocol with an explicit Learned-Fixed attribution control and a documented synthetic-stability qualification, enabling a clearer distinction between what is supported by the current evidence and what remains unresolved.
\end{enumerate}

The remainder of this paper is organized as follows. Section~\ref{sec:method} presents the methodology, including the safe conditioning formulation, the bottleneck-side Prompt-Adapter design, and the locked training and evaluation protocol. Section~\ref{sec:results} reports the experimental results, including controlled internal comparisons, attribution analysis, and stability qualifications. Section~\ref{sec:conclusion} concludes the paper and discusses current limitations and possible future directions.

\section{Methodology}
\label{sec:method}

\subsection{Problem Framing and Design Rationale}

Low-light image enhancement is not a simple brightness-adjustment task. A successful LLIE model must recover content under spatially non-uniform illumination while suppressing noise amplification, color distortion, and detail loss. Illumination priors are therefore relevant to the task, but the route by which the prior enters the restoration backbone is equally important. Exploratory runs in the present study indicated that direct or overly strong interaction with state-space dynamics can become numerically unstable and visually unsafe. The final design is built around this observation. Rather than maximizing interaction strength, the method seeks a safer conditioning path that preserves the restoration backbone while still allowing illumination guidance to influence the recovered image.

\subsection{Overall Pipeline}

\begin{figure*}[!tb]
\centering
\includegraphics[width=0.92\textwidth]{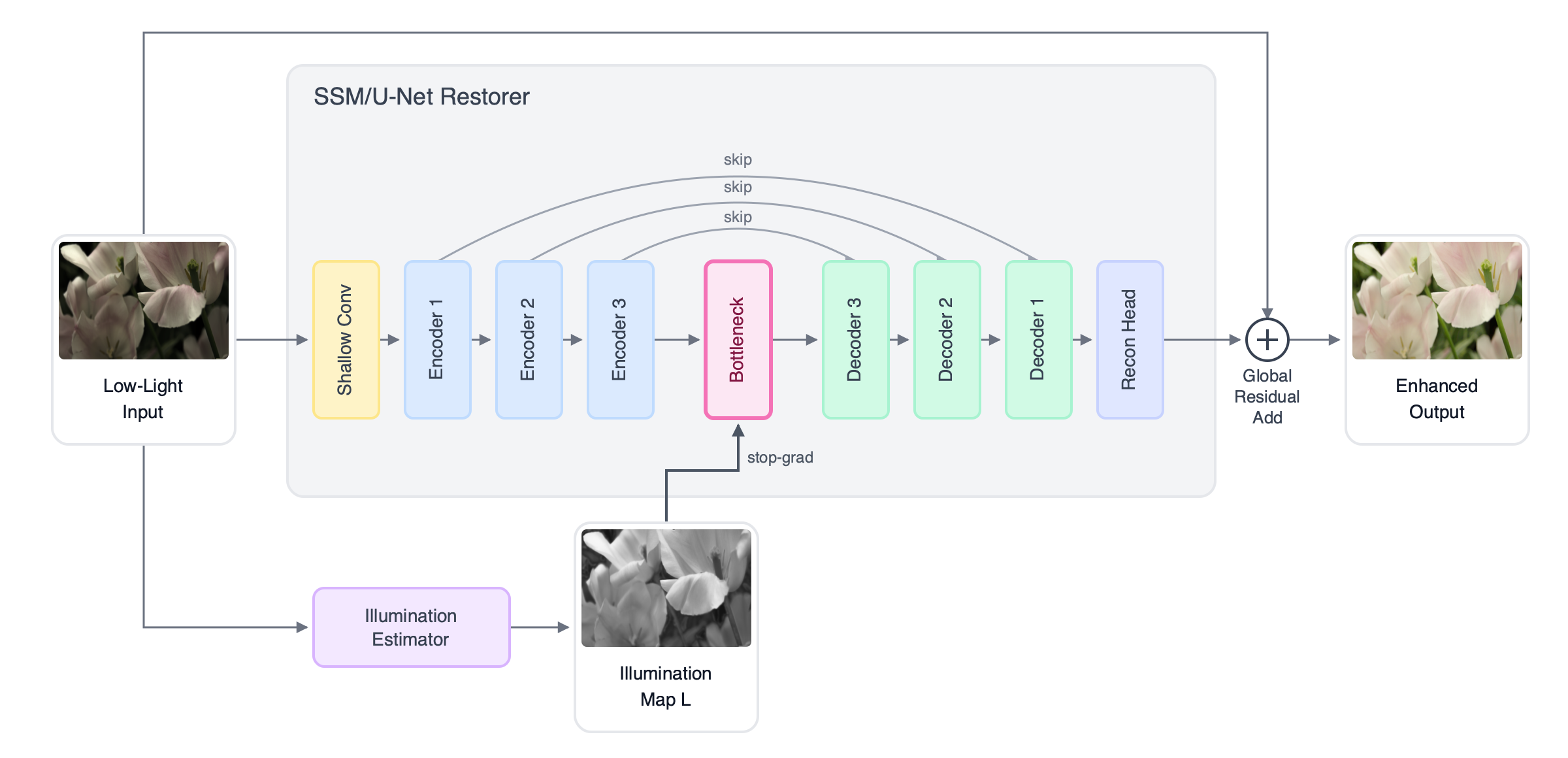}
\caption{Overall architecture of the proposed SSM-based LLIE restorer with bottleneck-side Prompt-Adapter.}
\label{fig:mamballie-architecture}
\end{figure*}

\begin{figure*}[!tb]
\centering
\includegraphics[width=0.88\textwidth]{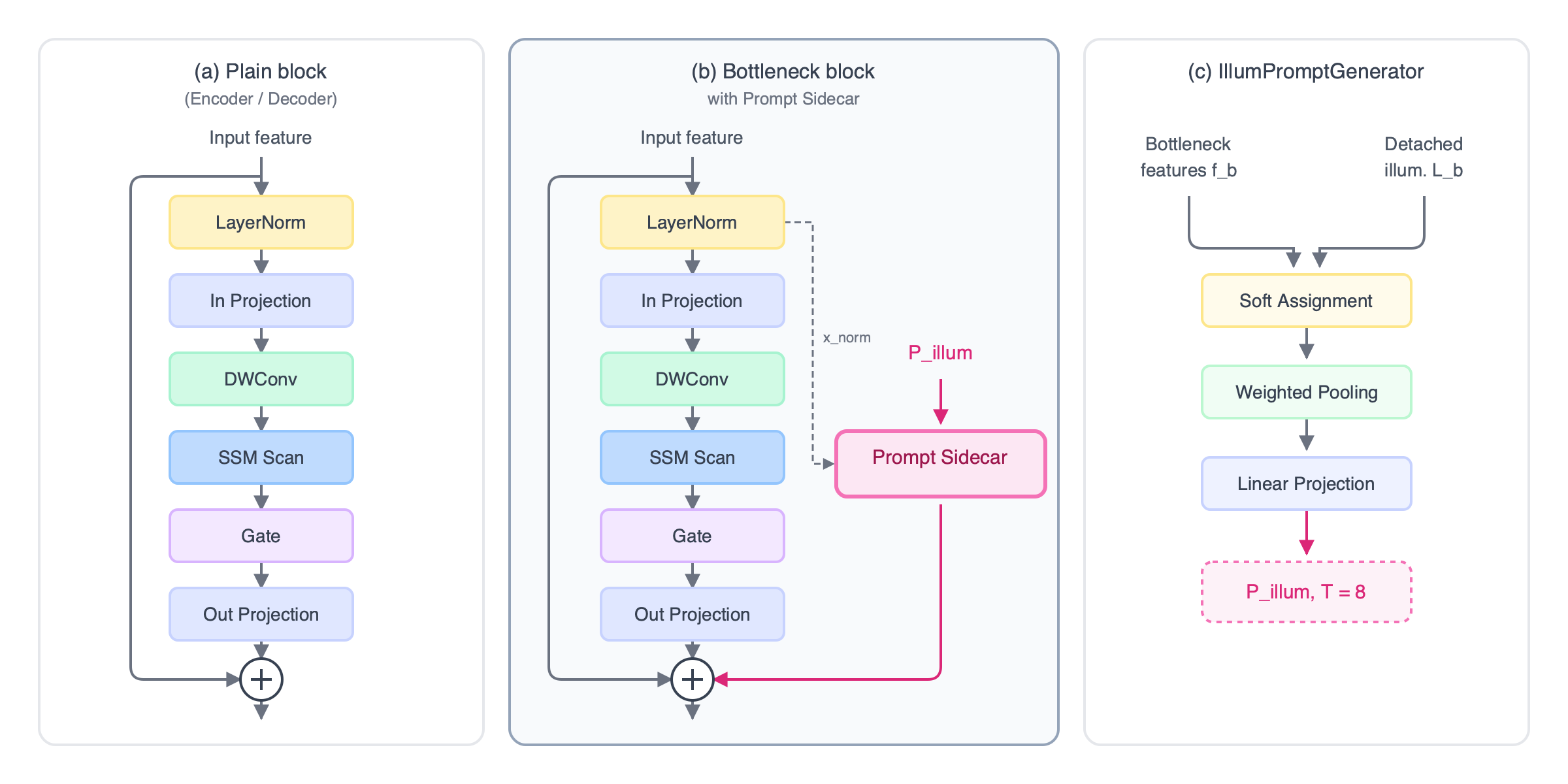}
\caption{Detail of the sidecar conditioning block with cross-attention prompt injection mechanism.}
\label{fig:mamballie-block}
\end{figure*}

The evaluated restorer follows a two-stage design, illustrated in Fig.~\ref{fig:mamballie-architecture} and Fig.~\ref{fig:mamballie-block}. Given a low-light input image $\mathbf{x} \in [0,1]^{3 \times H \times W}$, a lightweight illumination estimator first predicts an illumination map $\mathbf{L} \in [0,1]^{1 \times H \times W}$. The restorer then reconstructs an enhanced image $\hat{\mathbf{y}}$ using both the image features and the illumination signal:
\begin{equation}
\mathbf{L} = \mathrm{IE}(\mathbf{x}), \qquad
\hat{\mathbf{y}} = \mathrm{Restorer}(\mathbf{x}, \mathbf{L}).
\end{equation}

In the final main route, the illumination signal does not directly modify the core state-space transition. Instead, it is converted into prompts and injected through a bottleneck-side adapter. This sidecar design is the main architectural choice of the paper.

\subsection{Illumination Estimator}

The illumination estimator is a lightweight convolutional network consisting of three ReLU-activated $3 \times 3$ convolutions followed by a final $3 \times 3$ convolution with sigmoid activation, producing a single-channel map. The network is intentionally small relative to the restoration backbone because its role is to provide a spatial prior rather than to solve the enhancement problem on its own.

Training uses a dedicated warm-up stage for the illumination estimator. Let $\mathbf{x}$ denote the low-light input and $\mathbf{y}$ the ground truth. The warm-up target is the channel-mean ratio between input and target:
\begin{equation}
\mathbf{L}^{\ast} = \mathrm{clip}\!\left(\mathrm{mean}_{c}\left(\frac{\mathbf{x}}{\mathbf{y} + \epsilon}\right), 0, 1\right).
\end{equation}
The warm-up loss is then
\begin{equation}
\mathcal{L}_{\mathrm{IE}} = \|\mathbf{L} - \mathbf{L}^{\ast}\|_{2}^{2}.
\end{equation}
This stage gives the illumination branch a stable starting point before joint optimization with the restorer.

\subsection{Base SSM/U-Net Restorer}

The restoration backbone, denoted IAMRN in the implementation, is a three-scale U-Net style network built on selective state-space (Mamba) blocks~\cite{gu2023mamba} with channel widths $(48, 96, 192)$, two encoder-decoder blocks per scale, and two bottleneck blocks. A shallow convolution first extracts low-level features from the input image, after which the encoder progressively downsamples them to a bottleneck representation. The decoder mirrors this process and reconstructs the final RGB image with skip connections from the encoder.

The reported main runs disable IGN-based token reordering and keep the backbone matched across Base, Prompt-Adapter, and Learned-Fixed. This choice is deliberate: the main comparison is intended to isolate the conditioning route rather than to combine multiple structural changes into a single headline result.

\subsection{Illumination Prompt Generation}

At the bottleneck scale, the model converts feature maps and the illumination map into a small set of prompt tokens. Let $\mathbf{F} \in \mathbb{R}^{C \times H_b \times W_b}$ be bottleneck features and let $\mathbf{L}_b$ be the illumination map resized to the same resolution. The prompt generator first assigns each spatial location softly to one of $T$ illumination zones:
\begin{equation}
\mathbf{W}_{i} = \mathrm{softmax}(g(\mathbf{L}_{b,i})),
\end{equation}
where $g(\cdot)$ is a small multilayer perceptron and $\mathbf{W}_{i} \in \mathbb{R}^{T}$ is the zone-assignment distribution of location $i$. Features are then pooled within each zone and projected into the state dimension:
\begin{equation}
\mathbf{P}_{t} = h\!\left(\frac{\sum_{i} W_{i,t}\mathbf{F}_{i}}{\sum_{i} W_{i,t} + \delta}\right), \qquad t=1,\dots,T,
\end{equation}
where $h(\cdot)$ is a feature projection network. The result is a prompt bank $\mathbf{P} \in \mathbb{R}^{T \times d_{\mathrm{state}}}$ that summarizes illumination-aware side information at the bottleneck.

\subsection{Bottleneck-Side Prompt Adapter}

The key design choice is to inject prompts outside the SSM core rather than into its internal dynamics. Let $\mathbf{X}_{b}$ denote normalized bottleneck features reshaped into token form. The prompt sidecar adapter computes queries from $\mathbf{X}_{b}$, attends to the prompt bank, and projects the result back into the feature dimension:
\begin{gather}
\mathbf{Q} = \mathbf{X}_{b}\mathbf{W}_{q}, \qquad
\mathbf{A} = \mathrm{softmax}\!\left(\frac{\mathbf{Q}\mathbf{P}^{\top}}{\sqrt{d_{\mathrm{state}}}}\right)\mathbf{P}, \\
\mathbf{X}'_{b} = \mathbf{X}_{b} + s \cdot \phi(\mathbf{A}),
\end{gather}
where $\phi(\cdot)$ is the output projection and $s \in [0,1]$ is a module activation scale. In the implementation, the output projection is zero-initialized so that the sidecar branch starts as an identity residual. This conservative initialization reduces the risk of abruptly corrupting the restoration backbone at the beginning of joint training.

This bottleneck-only placement is deliberate. It injects global illumination-aware side information where the receptive field is largest while avoiding repeated interference with the full stack of state-space blocks. The method therefore treats conditioning as a controlled residual cue rather than as a full rewrite of the backbone dynamics.

\subsection{Learned-Fixed Attribution Control}

To separate illumination conditioning from prompt capacity, a Learned-Fixed control is introduced. This variant preserves the same sidecar adapter and prompt dimensionality as the main Prompt-Adapter, but replaces the input-dependent prompt generator with a learned prompt bank shared across samples:
\begin{equation}
\mathbf{P} = \mathbf{P}_{\mathrm{fixed}}.
\end{equation}
This control is not presented as a competing production model. Its role is interpretive: it tests whether the observed gain should be attributed to illumination-aware prompting itself or to the added sidecar capacity more generally.

\subsection{Training Objective and Stabilization}

After illumination-estimator warm-up, the full model is optimized with a composite objective:
\begin{equation}
\mathcal{L}_{\mathrm{total}}
= \mathcal{L}_{\mathrm{char}}
+ \mathcal{L}_{\mathrm{freq}}
+ \lambda_{p}\mathcal{L}_{\mathrm{percep}}
+ \alpha\lambda_{c}\mathcal{L}_{\mathrm{consist}}
+ \alpha\lambda_{s}\mathcal{L}_{\mathrm{smooth}}.
\end{equation}
Here $\mathcal{L}_{\mathrm{char}}$ is the Charbonnier reconstruction loss, $\mathcal{L}_{\mathrm{freq}}$ is a frequency-aware loss based on Laplacian responses, $\mathcal{L}_{\mathrm{percep}}$ is a VGG perceptual loss, and the last two terms are illumination consistency and smoothness losses. In the final reported main runs, $\lambda_{p}=0.01$ and the two illumination-regularization weights are set to zero, so the locked recipe emphasizes reconstruction, frequency preservation, and perceptual supervision while isolating the conditioning route itself.

Several stabilization measures are part of the method rather than incidental engineering. First, the illumination estimator is warmed up separately before joint training. Second, the illumination map can be detached before being passed into the restorer, which reduces unstable gradient coupling between the two stages; the final main runs use this detached setting. Third, the conditioning route is activated gradually through a module warm-up schedule rather than being applied at full strength from the first joint iteration. Finally, gradient clipping, mixed-precision scaler checkpointing, explicit seed control, and skip logging are retained in the training pipeline so that stability behavior is observable and reproducible.

\subsection{Implementation and Evaluation Settings}

The final main runs use AdamW optimization with a base learning rate of $3 \times 10^{-4}$, an illumination-estimator learning rate of $3 \times 10^{-5}$, weight decay $10^{-4}$, training patch size $128$, and batch size $8$. Phase~1 uses 5k illumination-estimator warm-up iterations. Phase~2 uses 150k joint iterations on LOL-v1 and 175k joint iterations on both LOLv2-Real and LOLv2-Synthetic. Cosine-restart scheduling, gradient clipping at 1.0, and AMP are enabled in the final recipe. Evaluation uses the locked Y-channel protocol, with $\tau=0.1$ and $\texttt{module\_scale}=1.0$ fixed at test time.

\begin{table}[t]
\centering
\small
\caption{Model complexity comparison across the three evaluated variants. Parameters are reported in millions (M) and FLOPs are estimated on a $128 \times 128$ input patch using fvcore.}
\label{tab:model-complexity}
\begin{tabular}{lcc}
\toprule
Variant & Params (M) & FLOPs (G) \\
\midrule
Base                & 2.46 & 5.16 \\
Prompt-Adapter      & 2.52 & 5.18 \\
Learned-Fixed       & 2.47 & 5.17 \\
\bottomrule
\end{tabular}
\end{table}

\section{Results}
\label{sec:results}

\subsection{Evaluation Protocol}

The proposed safe-conditioning route is evaluated on three paired low-light image enhancement benchmarks: LOL-v1~\cite{wei2018retinex}, LOLv2-Real, and LOLv2-Synthetic~\cite{yang2020fidelity}. LOL-v1 contains 485 training pairs and 15 testing pairs. LOLv2-Real contains 689 training pairs and 100 testing pairs captured from real low-light scenes, while LOLv2-Synthetic contains 900 training pairs and 100 testing pairs rendered from normally exposed images under a synthetic degradation process. These three datasets provide complementary test conditions: LOL-v1 is the standard small-scale paired benchmark, LOLv2-Real tests real captured low-light degradation, and LOLv2-Synthetic provides a broader paired synthetic setting.

The controlled internal evaluation uses the locked protocol defined for this study. PSNR and SSIM are computed on the Y channel under the BT.601 limited-range convention with $\texttt{data\_range}=255.0$, LPIPS~\cite{zhang2018unreasonable} is computed with the VGG backbone, and masked PSNR and masked SSIM are additionally reported on dark regions identified by the evaluation pipeline. These masked metrics are included because low-light enhancement should improve not only global fidelity but also reconstruction in the most severely underexposed image areas. All internal results are reported as mean $\pm$ standard deviation over three independent runs with seeds 3407, 3408, and 3409. At inference time, the conditioning hyper-parameters are fixed to $\tau=0.1$ and $\texttt{module\_scale}=1.0$, and no test-time augmentation is used.

Three closely matched variants are considered. \textbf{Base} removes the sidecar prompt pathway and measures the capability of the restoration backbone alone. \textbf{Prompt-Adapter} is the proposed route, in which illumination-conditioned prompts are injected through a bottleneck-side adapter rather than through direct modification of the state-space core. \textbf{Learned-Fixed} is an attribution control on LOL-v1 that preserves the same sidecar pathway and prompt capacity as Prompt-Adapter, but replaces the input-dependent prompt with a learned fixed prompt bank.

\subsection{Controlled Internal Comparison}

\begin{table*}[!t]
\centering
\small
\setlength{\tabcolsep}{4.2pt}
\renewcommand{\arraystretch}{1.05}
\caption{Controlled internal comparison under the locked three-seed Y-channel protocol (mean $\pm$ std over seeds 3407, 3408, 3409).}
\label{tab:internal-main}
\resizebox{\textwidth}{!}{%
\begin{tabular}{llccccc}
\toprule
Dataset & Variant & PSNR $\uparrow$ & SSIM $\uparrow$ & LPIPS $\downarrow$ & mPSNR $\uparrow$ & mSSIM $\uparrow$ \\
\midrule
LOL-v1 & Base & 25.01 $\pm$ 0.34 & 0.9077 $\pm$ 0.0011 & 0.2292 $\pm$ 0.0017 & 24.95 $\pm$ 0.21 & 0.9033 $\pm$ 0.0009 \\
LOL-v1 & Prompt-Adapter & 25.95 $\pm$ 0.18 & 0.9094 $\pm$ 0.0010 & \textbf{0.2273 $\pm$ 0.0066} & 25.96 $\pm$ 0.28 & 0.9053 $\pm$ 0.0006 \\
LOL-v1 & Learned-Fixed & \textbf{25.99 $\pm$ 0.59} & \textbf{0.9101 $\pm$ 0.0017} & 0.2289 $\pm$ 0.0063 & \textbf{26.09 $\pm$ 0.67} & \textbf{0.9063 $\pm$ 0.0021} \\
\midrule
LOLv2-Real & Base & 20.73 $\pm$ 0.47 & 0.8855 $\pm$ 0.0016 & 0.2415 $\pm$ 0.0056 & 20.99 $\pm$ 0.43 & 0.8818 $\pm$ 0.0018 \\
LOLv2-Real & Prompt-Adapter & \textbf{21.97 $\pm$ 0.34} & \textbf{0.8903 $\pm$ 0.0075} & \textbf{0.2309 $\pm$ 0.0153} & \textbf{22.42 $\pm$ 0.38} & \textbf{0.8876 $\pm$ 0.0077} \\
\midrule
LOLv2-Synthetic & Base & 25.44 $\pm$ 0.17 & 0.9634 $\pm$ 0.0007 & 0.0764 $\pm$ 0.0011 & 26.89 $\pm$ 0.40 & 0.9498 $\pm$ 0.0009 \\
LOLv2-Synthetic & Prompt-Adapter & \textbf{26.37 $\pm$ 0.38} & \textbf{0.9661 $\pm$ 0.0017} & \textbf{0.0710 $\pm$ 0.0006} & \textbf{27.80 $\pm$ 0.46} & \textbf{0.9517 $\pm$ 0.0029} \\
\bottomrule
\end{tabular}
}
\end{table*}

Table~\ref{tab:internal-main} reports the controlled internal comparison under the locked protocol. Across all three datasets, the Prompt-Adapter improves over the Base model. On LOL-v1, PSNR increases from $25.01 \pm 0.34$ dB to $25.95 \pm 0.18$ dB, LPIPS decreases from $0.2292 \pm 0.0017$ to $0.2273 \pm 0.0066$, and masked PSNR increases from $24.95 \pm 0.21$ dB to $25.96 \pm 0.28$ dB. The improvement therefore extends to the darker regions rather than being confined to already well-exposed areas.

The clearest real-data evidence appears on LOLv2-Real. The Prompt-Adapter improves PSNR from $20.73 \pm 0.47$ dB to $21.97 \pm 0.34$ dB, a gain of $+1.24$ dB. SSIM also increases from $0.8855 \pm 0.0016$ to $0.8903 \pm 0.0075$, while LPIPS decreases from $0.2415 \pm 0.0056$ to $0.2309 \pm 0.0153$. The masked PSNR gain is also large, increasing from $20.99 \pm 0.43$ dB to $22.42 \pm 0.38$ dB. Among the three evaluated datasets, this is the strongest evidence that the bottleneck-side prompt route improves restoration under realistic low-light degradation.

On LOLv2-Synthetic, the same direction holds: PSNR improves from $25.44 \pm 0.17$ dB to $26.37 \pm 0.38$ dB, SSIM improves from $0.9634 \pm 0.0007$ to $0.9661 \pm 0.0017$, LPIPS decreases from $0.0764 \pm 0.0011$ to $0.0710 \pm 0.0006$, and masked PSNR increases from $26.89 \pm 0.40$ dB to $27.80 \pm 0.46$ dB. This result is positive, but it must be read with a stability qualification discussed in Section~\ref{subsec:synthetic-stability}.

\subsection{Attribution Boundary With Learned-Fixed Prompts}

\begin{figure*}[!tb]
\centering
\setlength{\tabcolsep}{1pt}
\renewcommand{\arraystretch}{1.0}
\begin{tabular}{ccccc}
\textbf{Input} & \textbf{Base} & \textbf{Learned-Fixed} & \textbf{Prompt-Adapter} & \textbf{Ground Truth} \\
\includegraphics[width=0.182\textwidth]{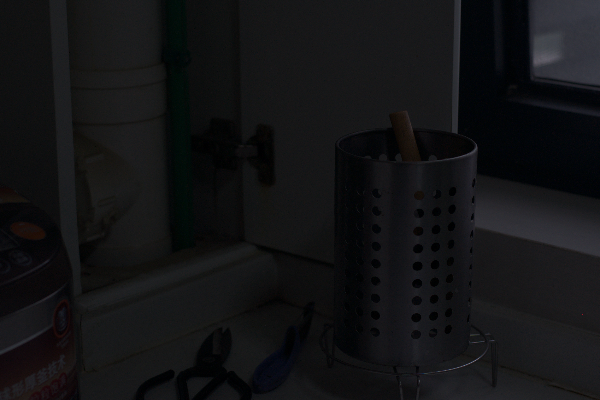} &
\includegraphics[width=0.182\textwidth]{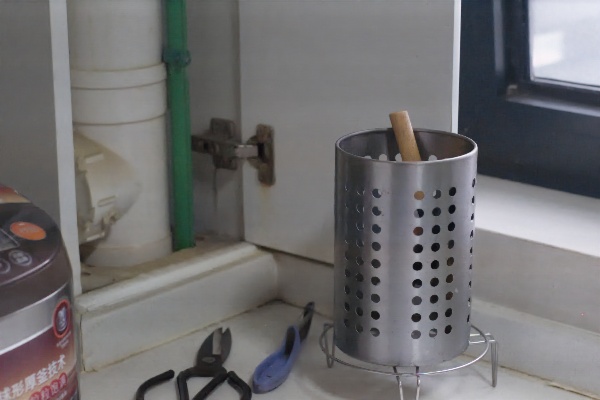} &
\includegraphics[width=0.182\textwidth]{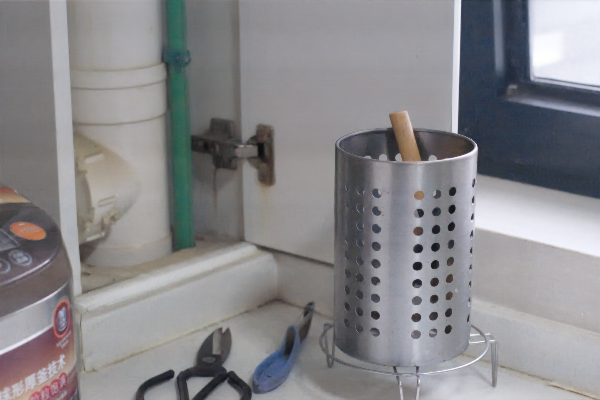} &
\includegraphics[width=0.182\textwidth]{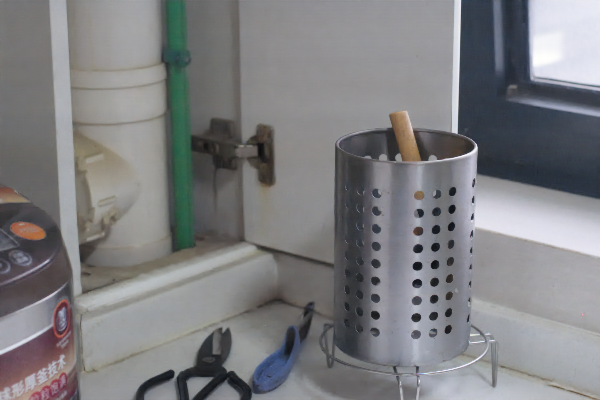} &
\includegraphics[width=0.182\textwidth]{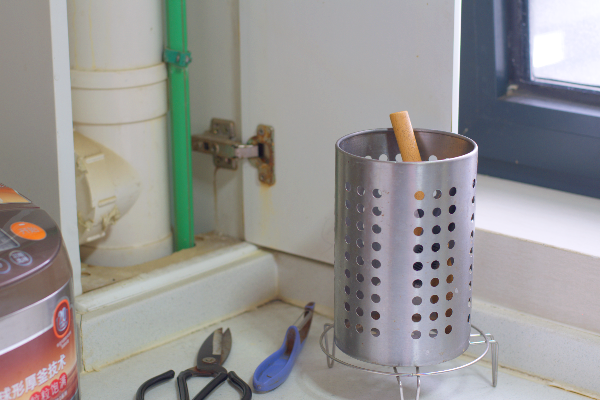} \\
\includegraphics[width=0.182\textwidth]{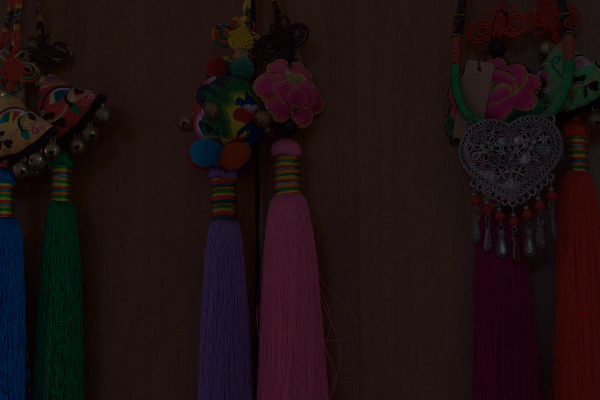} &
\includegraphics[width=0.182\textwidth]{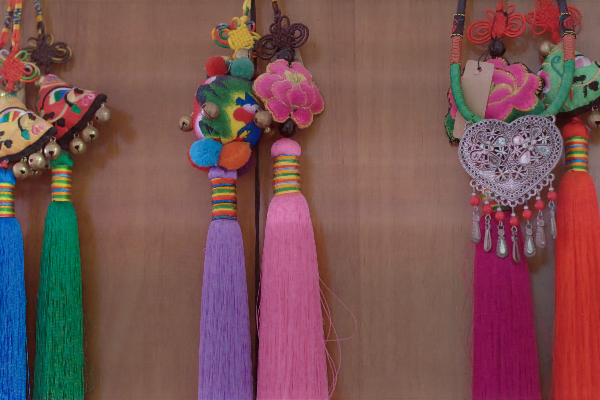} &
\includegraphics[width=0.182\textwidth]{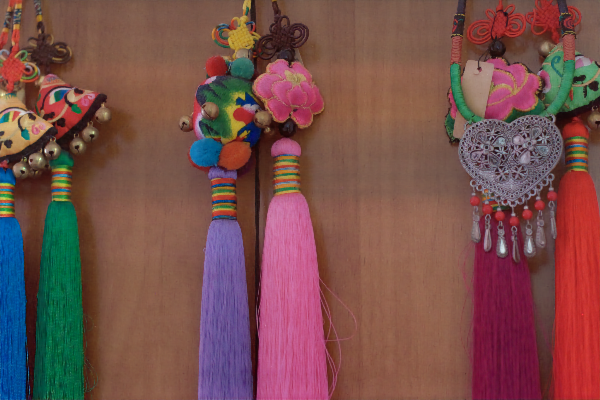} &
\includegraphics[width=0.182\textwidth]{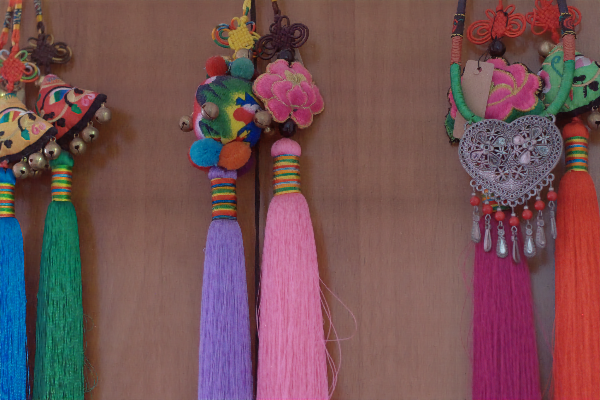} &
\includegraphics[width=0.182\textwidth]{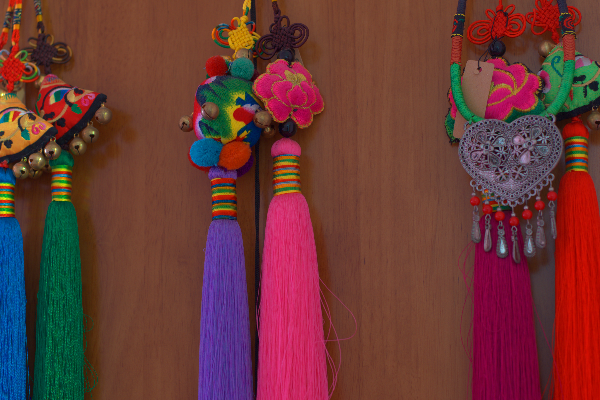} \\
\includegraphics[width=0.182\textwidth]{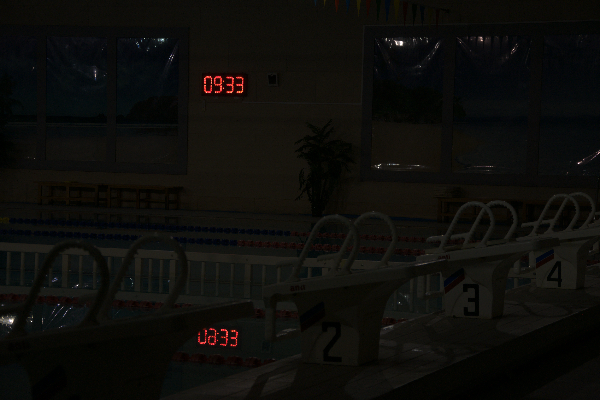} &
\includegraphics[width=0.182\textwidth]{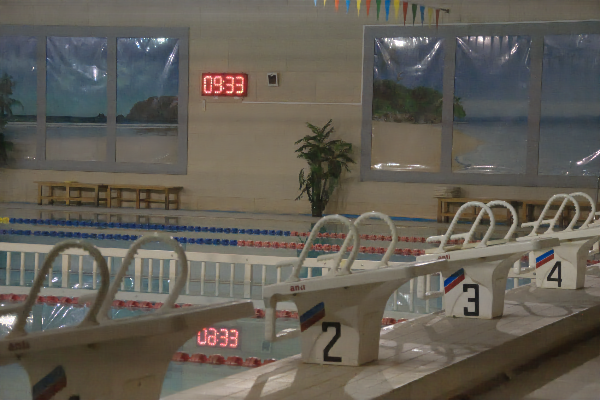} &
\includegraphics[width=0.182\textwidth]{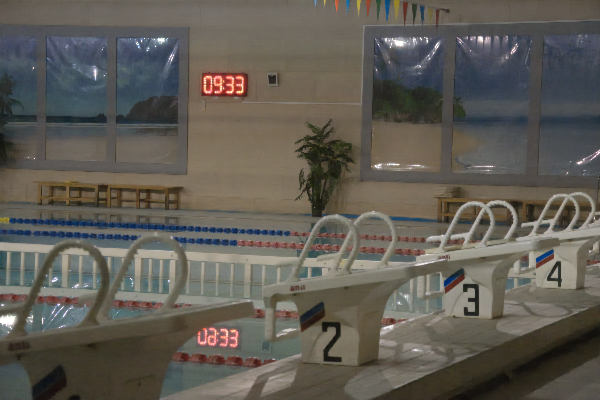} &
\includegraphics[width=0.182\textwidth]{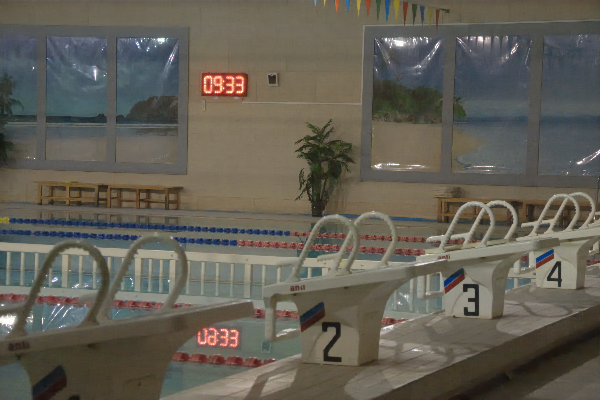} &
\includegraphics[width=0.182\textwidth]{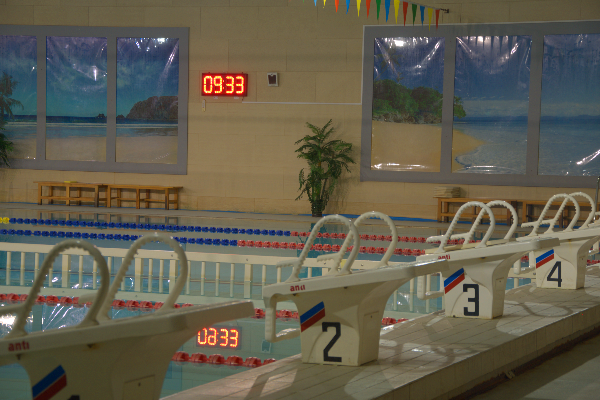} \\
\includegraphics[width=0.182\textwidth]{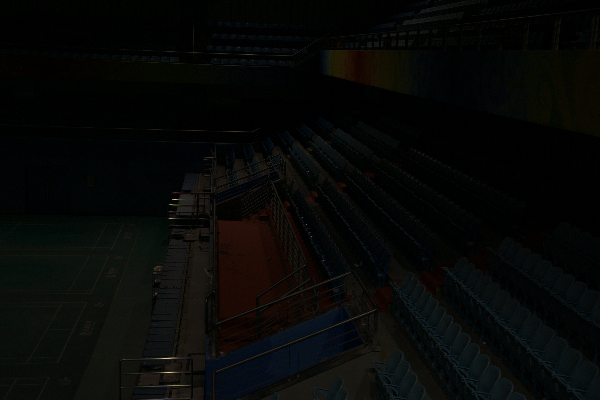} &
\includegraphics[width=0.182\textwidth]{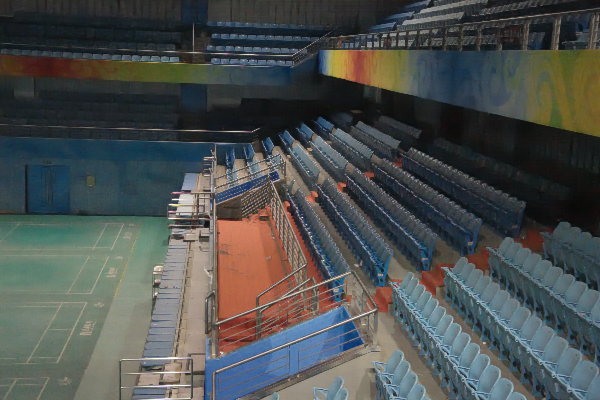} &
\includegraphics[width=0.182\textwidth]{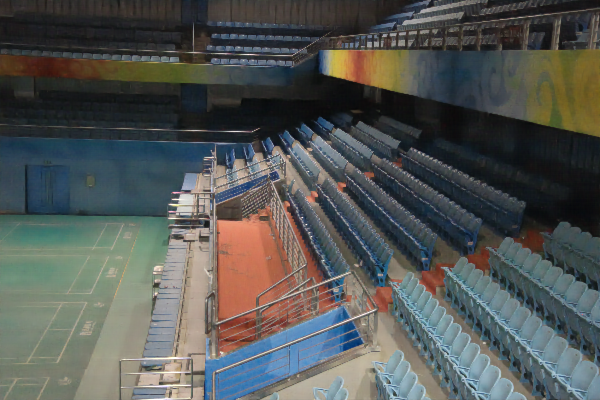} &
\includegraphics[width=0.182\textwidth]{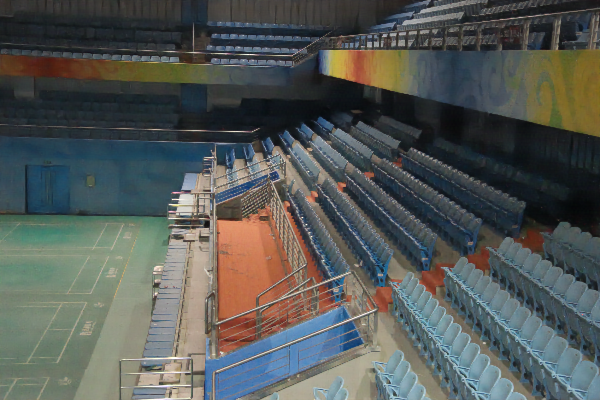} &
\includegraphics[width=0.182\textwidth]{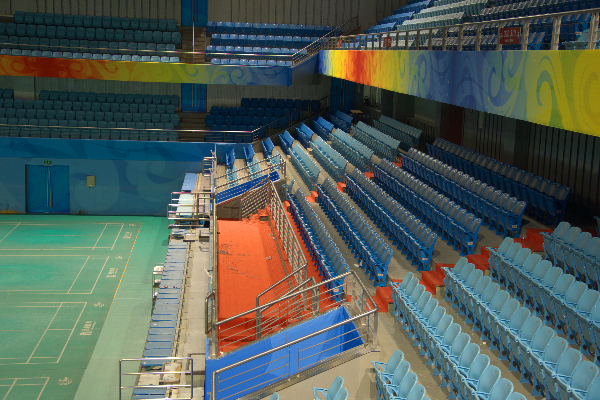} \\
\end{tabular}
\caption{Attribution-oriented qualitative comparison on LOL-v1 showing Base, Learned-Fixed, and Prompt-Adapter outputs against ground truth.}
\label{fig:qualitative-attribution}
\end{figure*}

A controlled gain over the Base model does not by itself prove that illumination-conditioned prompting is the unique cause of the improvement. The LOL-v1 Learned-Fixed control is therefore critical for interpretation. Under the locked protocol, Prompt-Adapter reaches $25.95 \pm 0.18$ dB PSNR, while Learned-Fixed reaches $25.99 \pm 0.59$ dB. The mean PSNR gap between the two variants is only $-0.04$ dB, so the present evidence does not support the stronger claim that illumination-conditioned prompts are strictly superior to learned fixed prompts in mean performance.

The more defensible conclusion is narrower. Base reaches $25.01 \pm 0.34$ dB, while both sidecar-prompt variants reach approximately $26$ dB, which indicates that the sidecar pathway itself is useful. At the same time, the Learned-Fixed control exhibits substantially larger variance: its PSNR standard deviation is $0.59$ dB compared with $0.18$ dB for Prompt-Adapter, and the same pattern appears in masked PSNR ($0.67$ dB versus $0.28$ dB). The current evidence therefore supports the usefulness of the sidecar conditioning route, while leaving illumination-specific superiority as an open question.

Figure~\ref{fig:qualitative-attribution} provides the matched qualitative view for this attribution boundary on the same four LOL-v1 samples. It should be read in the same bounded way as Table~\ref{tab:internal-main}: both sidecar-prompt variants visibly improve over the Base model on the selected examples, but the visual evidence does not justify a stronger claim of uniform Prompt-Adapter superiority over Learned-Fixed.

\subsection{Stability Qualification on LOLv2-Synthetic}
\label{subsec:synthetic-stability}

The LOLv2-Synthetic result should be reported with an explicit stability caveat. Earlier prompt-conditioned synthetic runs showed non-trivial skip counts under the mixed-precision training recipe, and the submission hardening process added scaler-state recovery, skip logging, and bad-batch tracing specifically to make this risk visible. In the tracked post-hardening runs, LOLv2-Synthetic Prompt-Adapter still shows materially larger skip counts than the corresponding Base runs, with documented counts of 1454 and 1692 on the new prompt-conditioned seeds, whereas the matching synthetic Base runs are clean in the same summary.

For this reason, LOLv2-Synthetic is not used as an unconstrained headline result. The correct interpretation is that the Prompt-Adapter improves over the Base model on this dataset under the final locked evaluation protocol, but the synthetic setting remains more fragile than the real-data evaluation. In contrast, LOLv2-Real does not carry the same stability burden in the current paper narrative and is therefore treated as the strongest cross-dataset evidence.

\subsection{Contextual Positioning Against Representative LLIE Methods}

\begin{table*}[!t]
\centering
\small
\setlength{\tabcolsep}{4.5pt}
\renewcommand{\arraystretch}{1.05}
\caption{Cross-protocol context for representative LLIE methods. External rows are cited from their primary sources under the original evaluation protocols; rows from this study use the Y-channel protocol of Section~\ref{sec:results}. The two groups are not directly comparable, and no ranking is implied.}
\label{tab:external-context}
\resizebox{\textwidth}{!}{%
\begin{tabular}{lccc p{2.35cm} p{2.55cm}}
\toprule
Method & LOL-v1 & LOLv2-Real & LOLv2-Syn & Relevance & Protocol note \\
\midrule
SNR-Net & 24.61 / 0.842 & 21.48 / 0.849 & 24.14 / 0.928 & degradation-aware CNN & PSNR/SSIM only \\
Retinexformer & 25.16 / 0.845 & 22.80 / 0.840 & 25.67 / 0.930 & illumination-guided Transformer & official ICCV 2023 table \\
RetinexMamba$^\dagger$ & 24.03 / 0.827 & 22.45 / 0.844 & 25.89 / 0.935 & SSM/Mamba-family reference & protocol footnote$^{\mathrm{a}}$ \\
Base (this study) & 25.01 / 0.9077 & 20.73 / 0.8855 & 25.44 / 0.9634 & reproduced base model & Y-channel (BT.601) \\
Prompt-Adapter (this study) & 25.95 / 0.9094 & 21.97 / 0.8903 & 26.37 / 0.9661 & reproduced safe-conditioning route & Y-channel (BT.601); synthetic caveat \\
\bottomrule
\end{tabular}
}

\vspace{2pt}
{\raggedright\footnotesize $^{\mathrm{a}}$RetinexMamba values are rounded to two decimal places from the original three-decimal-place figures reported in the primary source.\par}
\end{table*}

To situate the current study without turning the section into a broad benchmark catalog, Table~\ref{tab:external-context} compares the reproduced variants from this study with three representative citation baselines that are directly relevant to the present theme. SNR-Net~\cite{xu2022snr} represents a strong degradation-aware CNN line, Retinexformer~\cite{cai2023retinexformer} represents a modern illumination-guided Transformer line, and RetinexMamba~\cite{bai2024retinexmamba} provides a published SSM/Mamba-family reference. These rows are used only as citation context. They are not rerun under the internal protocol and therefore do not serve as the basis of the paper's causal claim.

The comparison is cross-protocol and must be read as such. The internal rows report PSNR and SSIM on the Y channel under the BT.601 convention, whereas the cited rows follow the evaluation conventions of their original papers; Retinexformer, for example, reports metrics in RGB space. Y-channel and RGB metrics are not numerically interchangeable, and the difference is already visible in the SSIM column, where the internal values and the cited values occupy different ranges. The table therefore supports no ordering between the methods of this study and the cited methods in either direction. The only quantitative comparison it contains that is protocol-matched is the Base-to-Prompt-Adapter gain, which is already reported in Table~\ref{tab:internal-main}. Establishing a ranking against published methods would require rerunning them under the same protocol, which is listed as a limitation in Section~\ref{sec:conclusion}.

\subsection{Qualitative Analysis}

\begin{figure*}[!tb]
\centering
\setlength{\tabcolsep}{2pt}
\renewcommand{\arraystretch}{1.0}
\begin{tabular}{c c c c}
\textbf{Input} & \textbf{Base} & \textbf{Prompt-Adapter} & \textbf{Ground Truth} \\[2pt]
\includegraphics[width=0.2\textwidth]{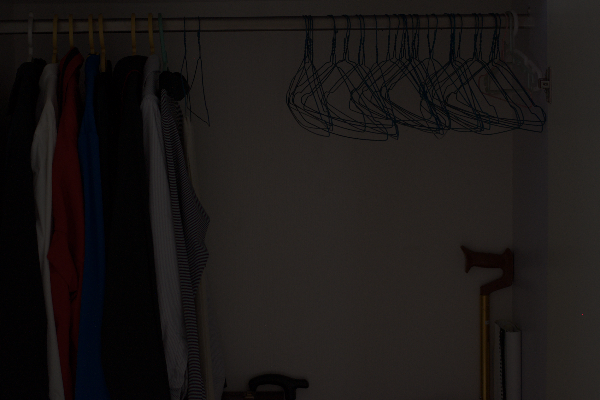} &
\includegraphics[width=0.2\textwidth]{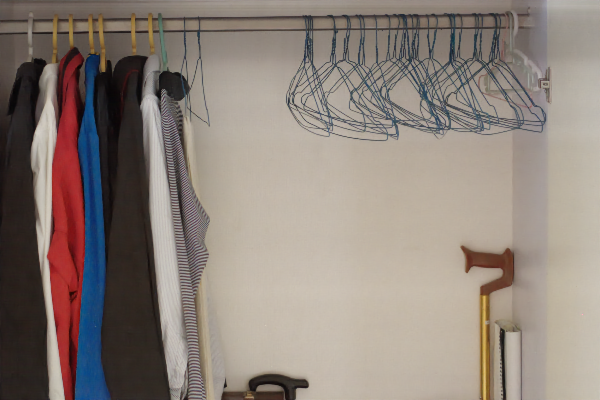} &
\includegraphics[width=0.2\textwidth]{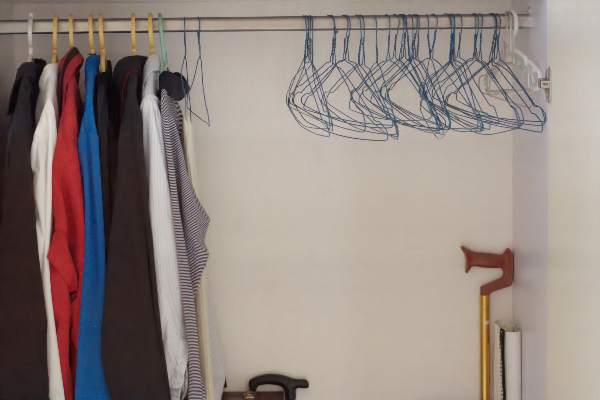} &
\includegraphics[width=0.2\textwidth]{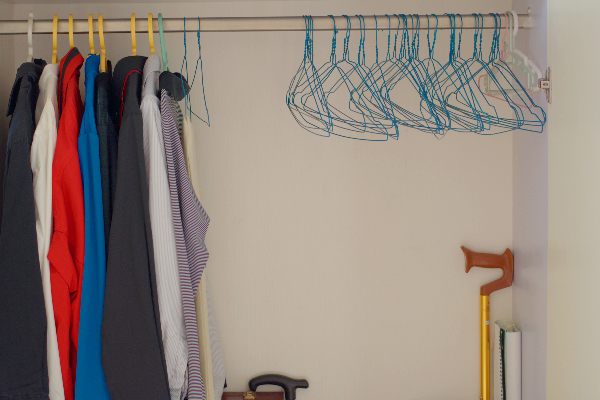} \\
\includegraphics[width=0.2\textwidth]{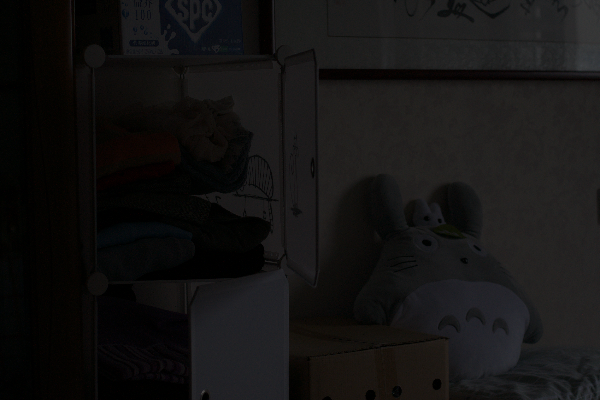} &
\includegraphics[width=0.2\textwidth]{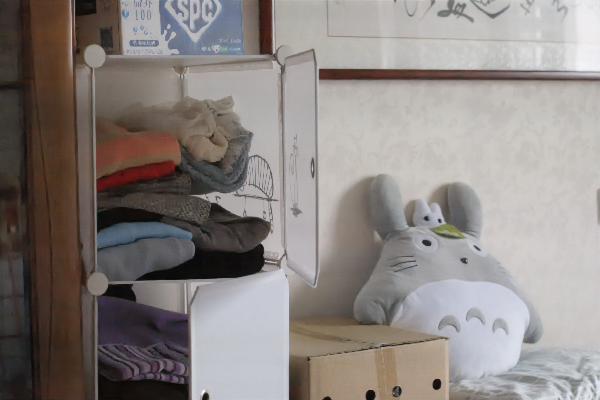} &
\includegraphics[width=0.2\textwidth]{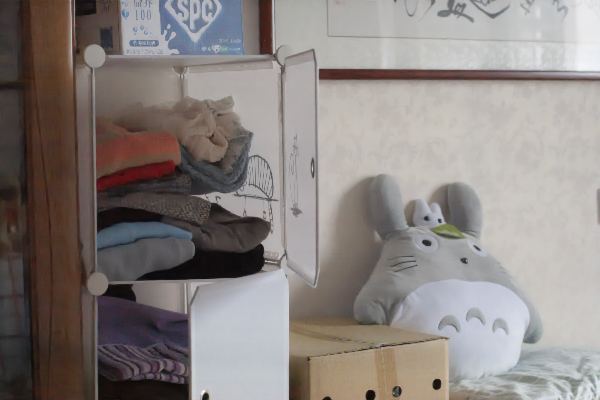} &
\includegraphics[width=0.2\textwidth]{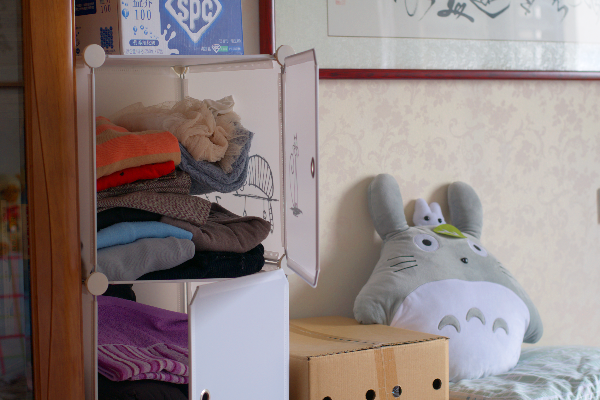} \\
\includegraphics[width=0.2\textwidth]{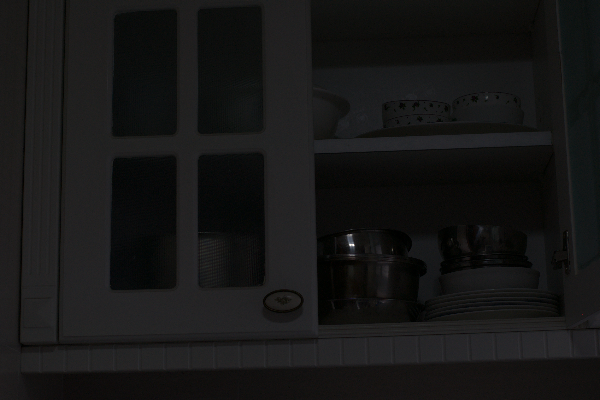} &
\includegraphics[width=0.2\textwidth]{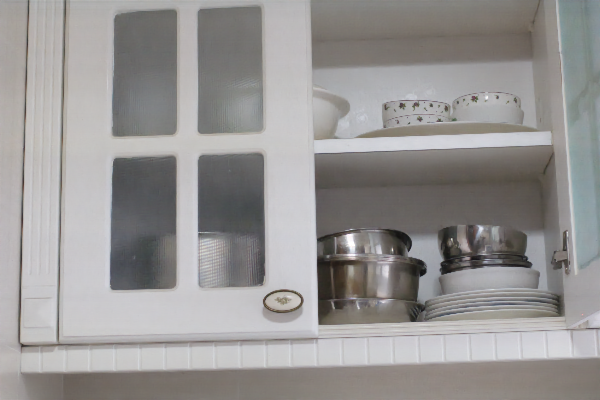} &
\includegraphics[width=0.2\textwidth]{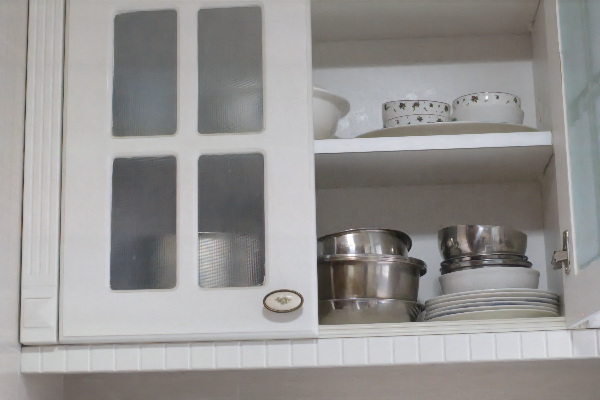} &
\includegraphics[width=0.2\textwidth]{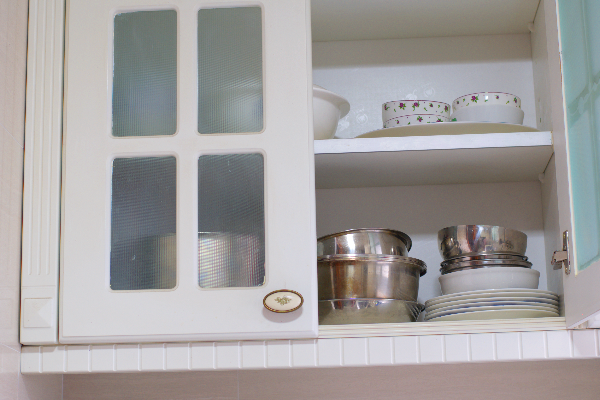} \\
\includegraphics[width=0.2\textwidth]{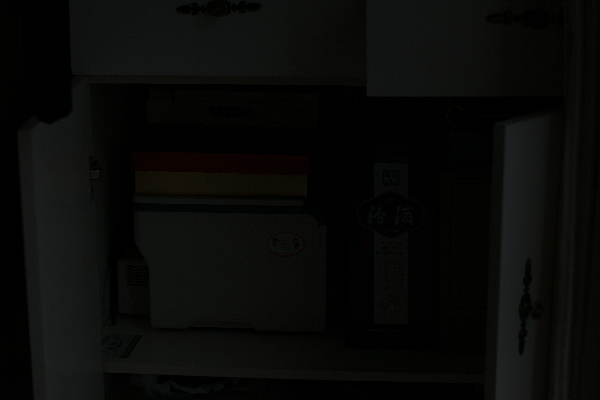} &
\includegraphics[width=0.2\textwidth]{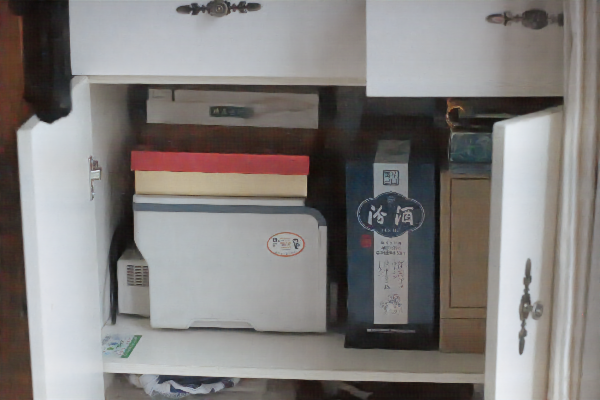} &
\includegraphics[width=0.2\textwidth]{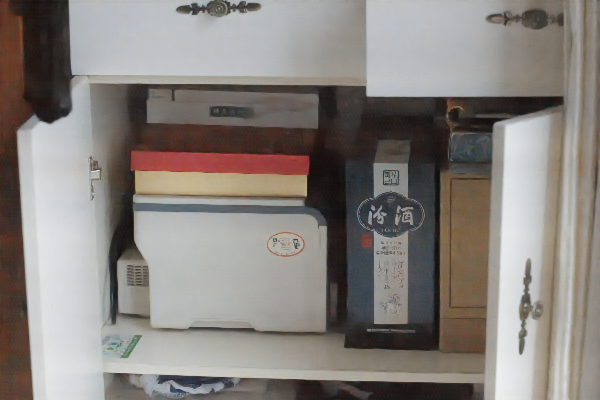} &
\includegraphics[width=0.2\textwidth]{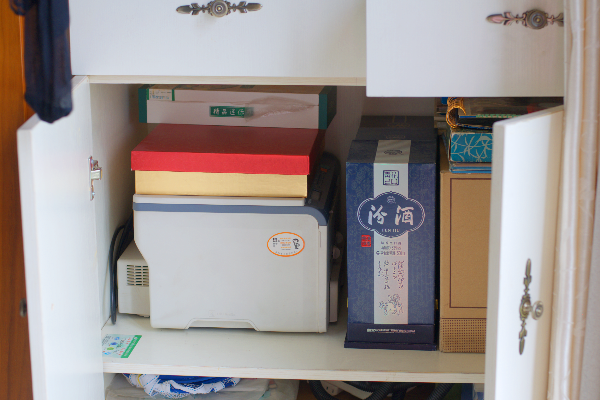} \\
\includegraphics[width=0.2\textwidth]{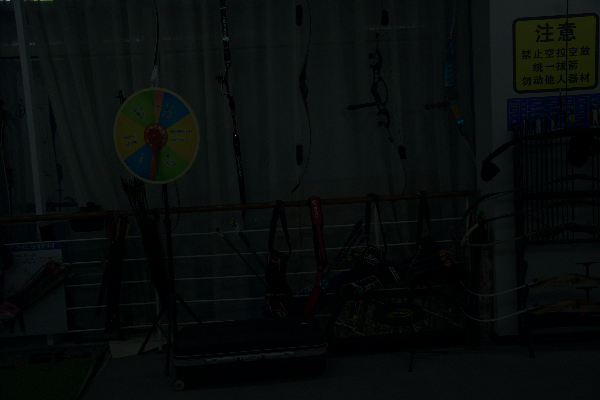} &
\includegraphics[width=0.2\textwidth]{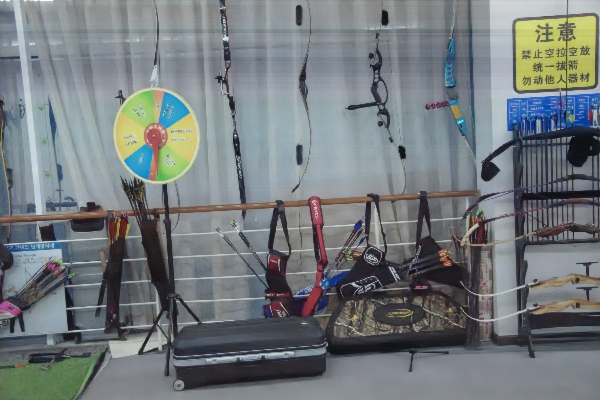} &
\includegraphics[width=0.2\textwidth]{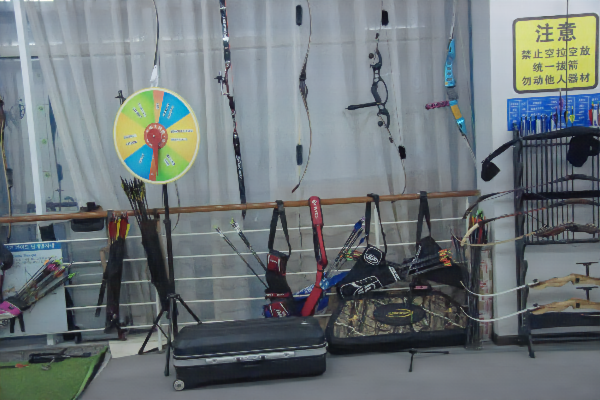} &
\includegraphics[width=0.2\textwidth]{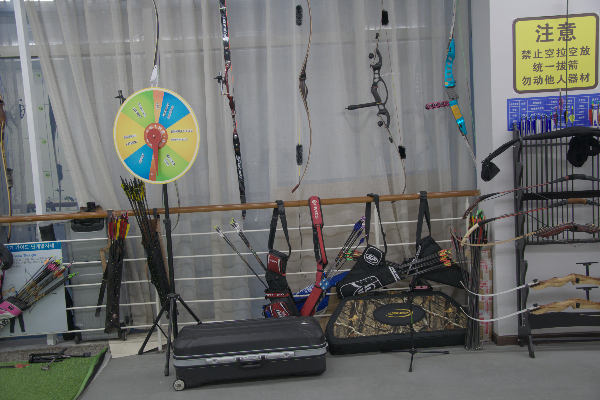} \\
\includegraphics[width=0.2\textwidth]{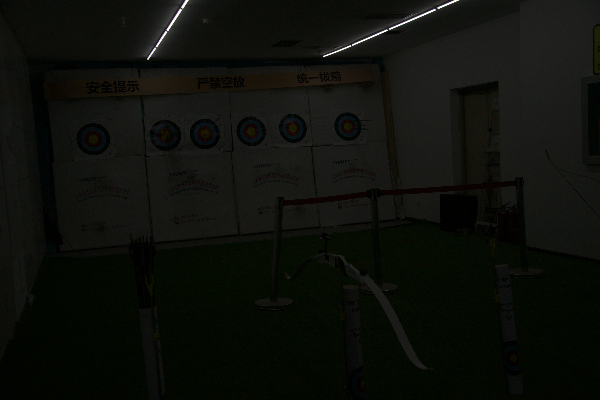} &
\includegraphics[width=0.2\textwidth]{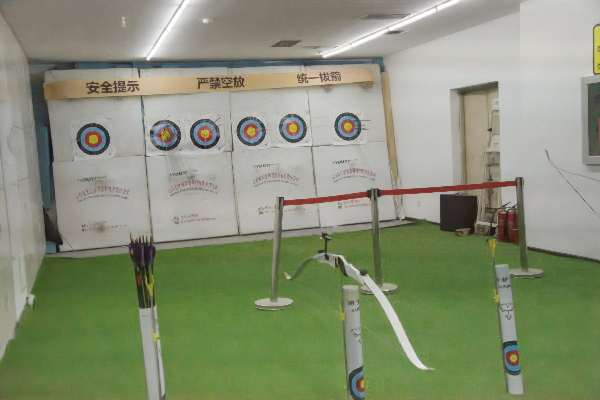} &
\includegraphics[width=0.2\textwidth]{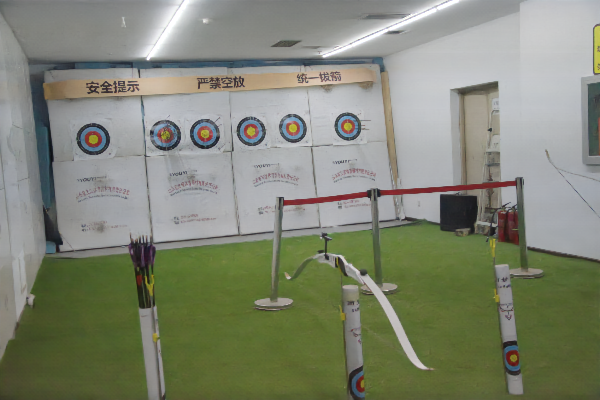} &
\includegraphics[width=0.2\textwidth]{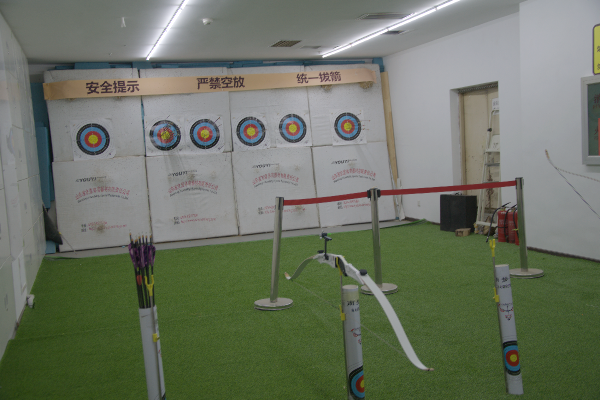} \\
\includegraphics[width=0.2\textwidth]{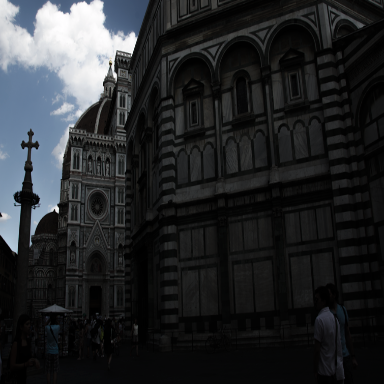} &
\includegraphics[width=0.2\textwidth]{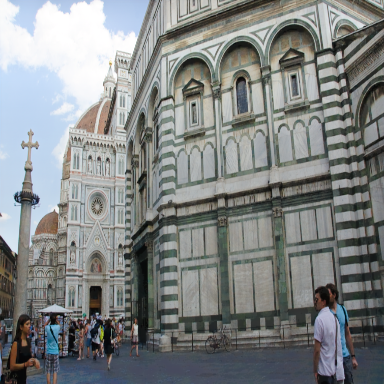} &
\includegraphics[width=0.2\textwidth]{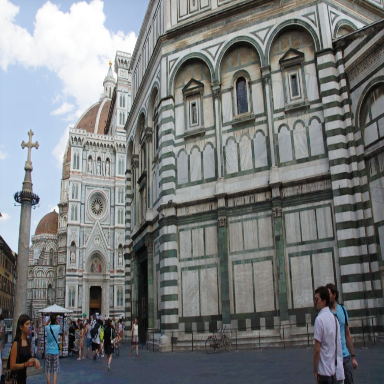} &
\includegraphics[width=0.2\textwidth]{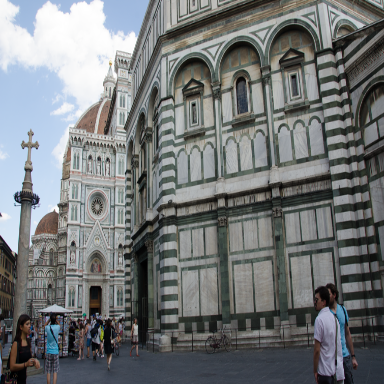} \\
\includegraphics[width=0.2\textwidth]{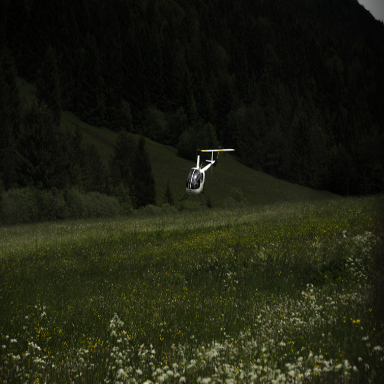} &
\includegraphics[width=0.2\textwidth]{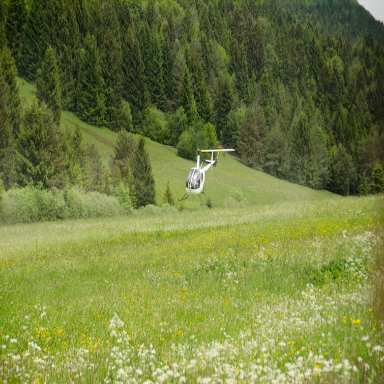} &
\includegraphics[width=0.2\textwidth]{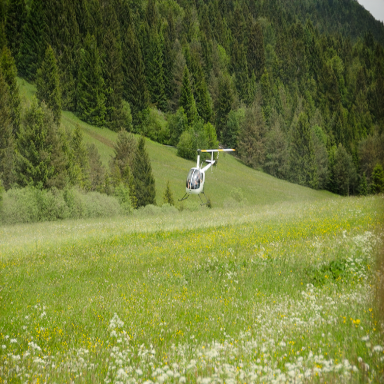} &
\includegraphics[width=0.2\textwidth]{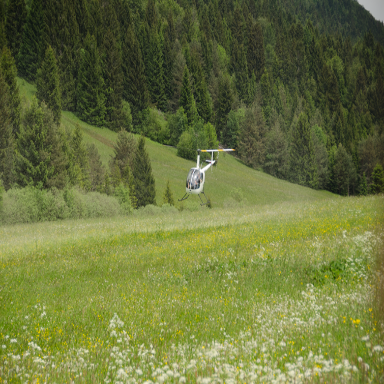} \\
\end{tabular}
\caption{Qualitative comparison between Base and Prompt-Adapter across LOL-v1, LOLv2-Real, and LOLv2-Synthetic test samples.}
\label{fig:qualitative-main}
\end{figure*}

Figure~\ref{fig:qualitative-main} presents representative samples from all three datasets, using a different set of LOL-v1 samples from the attribution panel in Fig.~\ref{fig:qualitative-attribution} to broaden the visual evidence. The qualitative comparison verifies that the Base-to-Prompt-Adapter gains correspond to visible changes in dark-region reconstruction, color fidelity, and artifact suppression.

In the LOL-v1 rows, the Prompt-Adapter shows improved color fidelity, finer texture recovery, more uniform brightening, and better dark-region structural preservation compared to the Base model. The LOLv2-Real rows show substantially improved brightness and detail recovery, consistent with the strongest quantitative gains in Table~\ref{tab:internal-main}. The LOLv2-Synthetic rows are included as bounded visual support and should be read with the stability caveat noted in Section~\ref{subsec:synthetic-stability}.

\subsection{Bounded Summary}

In summary, the Prompt-Adapter consistently improves over the Base model across all three datasets under the locked protocol, with LOLv2-Real providing the strongest real-data support. The Learned-Fixed control and the LOLv2-Synthetic stability caveat impose explicit boundaries on what the current evidence can claim.

\section{Limitations and Future Work}
\label{sec:conclusion}

Three limitations remain explicit. First, illumination-specific attribution is not fully isolated beyond the current LOL-v1 control, and extending comparable controls to LOLv2-Real would strengthen the causal interpretation of the method. Second, the optimization behavior on LOLv2-Synthetic remains more fragile than in the corresponding real-data setting, which motivates future work on more stable conditioning and training schemes. Third, the external comparison is still contextual rather than fully reproduced under a matched protocol. Future work should therefore prioritize stronger attribution controls, improved synthetic stability, and broader protocol-matched baseline evaluation rather than additional module accumulation.

\section{Conclusion}
This study examines safe illumination conditioning for SSM-based low-light image enhancement. The central issue is not simply whether illumination information is beneficial, but how that information should be introduced into an SSM-based restorer without disturbing the restoration dynamics. The current evidence supports a bottleneck-side Prompt-Adapter route as a practical answer to this question. By converting the estimated illumination signal into prompt tokens and injecting them through a residual sidecar pathway at the bottleneck, the method preserves the main restorer while consistently improving over the Base model under the locked evaluation protocol on LOL-v1, LOLv2-Real, and LOLv2-Synthetic.

The contribution of the study is intentionally bounded by the available evidence. LOLv2-Real provides the strongest real-data support for the proposed route, while the LOL-v1 Learned-Fixed control shows that the present results support the usefulness of the sidecar pathway more strongly than a strict illumination-specific superiority claim. The LOLv2-Synthetic result is also positive, but it remains stability-conditioned and should not be interpreted as a fully closed synthetic result. External results are reported only as cross-protocol context and do not establish a ranking against published methods.

\begingroup
\setlength{\parindent}{0pt}

\section*{Acknowledgement}
The authors would like to extend heartfelt gratitude to Xiamen University Malaysia for supporting this study under XMUM Research Fund: XMUMRF/2022-C10/IECE/0040.

\section*{Author Contribution Statement}
\textit{Wei Chen Ng}: Formal analysis, Investigation, Methodology, Writing -- original draft, Writing -- review \& editing.
\textit{Qiyi Xie}: Investigation, Methodology, Writing -- review \& editing.
\textit{Jian Ding Tan}: Conceptualization, Investigation, Funding acquisition, Supervision, Writing -- review \& editing.
\textit{Yan Kang Lee}: Validation, Writing -- review \& editing.
\textit{Abdulrab Hakim Qaid Abdullah}: Supervision, Writing -- review \& editing.
\textit{Wei Hown Tee}: Supervision, Writing -- review \& editing.

\section*{Declaration of Interests}
The authors declare that they have no known competing financial interests or personal relationships that could have appeared to influence the work reported in this paper.

\endgroup

\section*{Data Availability}
The source code, pretrained models, and evaluation scripts that support the findings of this study are openly available on GitHub at \url{https://github.com/WeiChenNg/SIC-BPS} and permanently archived on Zenodo at \url{https://doi.org/10.5281/zenodo.20588452}. The LOL-v1 and LOL-v2 benchmark datasets used in this study are publicly available from their original sources and are not redistributed here.

\bibliographystyle{IEEEtran}
\bibliography{references_main}

\appendices
\section{Per-Seed Results}
\label{app:per-seed}
Table~\ref{tab:appendix-per-seed} lists the seed-level results underlying the mean $\pm$ standard deviation values in Table~\ref{tab:internal-main}.

\begin{table*}[t]
\centering
\scriptsize
\setlength{\tabcolsep}{4.0pt}
\renewcommand{\arraystretch}{1.05}
\caption{Per-seed results underlying the locked internal comparison. All rows are evaluated under the locked Y-channel protocol with VGG-based LPIPS. This table provides the raw seed-level evidence for the mean $\pm$ standard deviation results reported in Table~\ref{tab:internal-main}.}
\label{tab:appendix-per-seed}
\resizebox{\textwidth}{!}{%
\begin{tabular}{llrccccc}
\toprule
Dataset & Variant & Seed & PSNR $\uparrow$ & SSIM $\uparrow$ & LPIPS $\downarrow$ & mPSNR $\uparrow$ & mSSIM $\uparrow$ \\
\midrule
LOL-v1 & Base & 3407 & 25.2991 & 0.9081 & 0.2288 & 25.1429 & 0.9035 \\
LOL-v1 & Base & 3408 & 24.6306 & 0.9065 & 0.2310 & 24.7266 & 0.9023 \\
LOL-v1 & Base & 3409 & 25.1019 & 0.9086 & 0.2277 & 24.9871 & 0.9041 \\
LOL-v1 & Prompt-Adapter & 3407 & 25.9464 & 0.9098 & 0.2218 & 25.9856 & 0.9059 \\
LOL-v1 & Prompt-Adapter & 3408 & 26.1289 & 0.9100 & 0.2346 & 26.2300 & 0.9052 \\
LOL-v1 & Prompt-Adapter & 3409 & 25.7692 & 0.9083 & 0.2255 & 25.6722 & 0.9047 \\
LOL-v1 & Learned-Fixed & 3407 & 25.4026 & 0.9082 & 0.2322 & 25.3555 & 0.9039 \\
LOL-v1 & Learned-Fixed & 3408 & 25.9844 & 0.9106 & 0.2216 & 26.2374 & 0.9073 \\
LOL-v1 & Learned-Fixed & 3409 & 26.5811 & 0.9116 & 0.2328 & 26.6654 & 0.9078 \\
LOLv2-Real & Base & 3407 & 20.5158 & 0.8872 & 0.2351 & 20.8819 & 0.8839 \\
LOLv2-Real & Base & 3408 & 20.4067 & 0.8854 & 0.2437 & 20.6190 & 0.8809 \\
LOLv2-Real & Base & 3409 & 21.2720 & 0.8840 & 0.2457 & 21.4648 & 0.8806 \\
LOLv2-Real & Prompt-Adapter & 3407 & 21.6154 & 0.8926 & 0.2183 & 22.0038 & 0.8896 \\
LOLv2-Real & Prompt-Adapter & 3408 & 22.2925 & 0.8963 & 0.2265 & 22.7608 & 0.8941 \\
LOLv2-Real & Prompt-Adapter & 3409 & 21.9972 & 0.8818 & 0.2480 & 22.4969 & 0.8791 \\
LOLv2-Synthetic & Base & 3407 & 25.2789 & 0.9629 & 0.0755 & 26.4335 & 0.9488 \\
LOLv2-Synthetic & Base & 3408 & 25.6154 & 0.9642 & 0.0776 & 27.1886 & 0.9506 \\
LOLv2-Synthetic & Base & 3409 & 25.4203 & 0.9633 & 0.0761 & 27.0406 & 0.9499 \\
LOLv2-Synthetic & Prompt-Adapter & 3407 & 26.8065 & 0.9679 & 0.0703 & 28.1985 & 0.9549 \\
LOLv2-Synthetic & Prompt-Adapter & 3408 & 26.1301 & 0.9659 & 0.0715 & 27.2962 & 0.9494 \\
LOLv2-Synthetic & Prompt-Adapter & 3409 & 26.1832 & 0.9646 & 0.0712 & 27.9033 & 0.9506 \\
\bottomrule
\end{tabular}
}
\end{table*}

\end{document}